\documentclass{llncs}

\usepackage{graphicx}
\usepackage[utf8]{inputenc}
\usepackage{amsmath,amsfonts,amssymb} 
\usepackage[T1]{fontenc}
\usepackage{graphicx} 
\usepackage{url}
\usepackage{placeins}
\usepackage{subfigure}

\begin{document}

\title{Feature extraction from complex networks: A case of study in genomic sequences classification}
\author{Bruno Mendes Moro Conque\inst{1}, Andr\'{e}  Yoshiaki Kashiwabara\inst{1}, Fabr\'{i}cio Martins Lopes\inst{1}}
\institute{Federal University of Technology - Paran\'{a}, Brazil\\
\email{fabricio@utfpr.edu.br}}

  
\maketitle

\begin{abstract}
\vspace{-0.3cm}
This work presents a new approach for classification of genomic sequences from measurements of complex networks and information theory.
For this, it is considered the nucleotides, dinucleotides and trinucleotides of a genomic sequence. 
For each of them, the entropy, sum entropy and maximum entropy values are calculated.
For each of them is also generated a network, in which the nodes are the nucleotides, dinucleotides or trinucleotides and
its edges are estimated by observing the respective adjacency among them in the genomic sequence.
In this way, it is generated three networks, for which measures of complex networks are extracted.
These measures together with measures of information theory comprise a feature vector representing a genomic sequence.
Thus, the feature vector is used for classification by methods such as \textsl{SVM}, \textsl{MultiLayer Perceptron}, 
\textsl{J48}, \textsl{IBK}, \textsl{Naive Bayes} and \textsl{Random Forest} in order to evaluate the proposed approach.
It was adopted coding sequences, intergenic sequences and TSS (Transcriptional Starter Sites) as datasets,
for which the better results were obtained by the \textsl{Random Forest} with 91.2\%, 
followed by \textsl{J48} with 89.1\% and \textsl{SVM} with 84.8\% of accuracy.
These results indicate that the new approach of feature extraction has its value, 
reaching good levels of classification even considering only the genomic sequences, i.e., 
no other a priori knowledge about them is considered.
\end{abstract}

\section{Introduction}\label{sec:intro}
\vspace{-0.1cm}
The study of biological systems and how its components are interconnected is a great challenge nowadays, 
attracting the attention of researchers from various fields of knowledge. 
This new scientific research area is known as systems biology and it is highly interdisciplinary.
The main focus is to analyze the organism in a holistic way.

The complex networks theory has been successfully applied in systems biology for modeling 
interactions between cellular components \cite{shen2002network,diambra2005complex,lopes2014feature,lopes2011gene}, 
proteins \cite{jeong2001lethality,costa2006protein} and metabolic relationships \cite{jeong2000large}.
Most often, the study of a interconnected components subset is more important than the individual analysis of them, 
which leads to a suitable application of complex networks in biology. As suggested by \cite{vogelstein2000surfing},
the p53 gene is a tumor suppressor, however it is more important to analyze its connections than studying it individually.

Through complex networks it is possible to extract measures that represent characteristics in natural and artificial 
systems composed of interacting parts.
However, despite the great success obtained by the complex network theory, 
there is still a huge field of research to be explored due to some limitations such as the lack of measures and 
methods to analyze, characterize and classify real networks.
Therefore, to obtain a more precise characterization, it is essential to consider a broad set of non-redundant 
measurements, which can be achieved with the use of pattern recognition techniques and data mining \cite{costa2007characterization}.
On the other hand, the information theory is also an important measure generally used for indicate the amount of information
of a given source, it was successfully applied in a myriad of bioinformatics problems in order to characterize the relationships
among genes \cite{butte2000,lopes2008b,lopes2011tsallis,lopes2014b}.
%

In the context of pattern recognition, the feature extraction is a form of dimensionality reduction \cite{theodoridis2008}.
More specifically, in a reductionist point of view, a feature extraction method attempts to represent a sample 
for a relatively small subset of features, while minimizing the loss of information from this sample.
As a result, if the feature extractor succeeds, the extracted features could be suitable to distinguish the samples 
even when considered samples belonging to different categories or classes.

Therefore, it is proposed in this work a new feature extraction approach for genomic sequences from complex networks and information theory.
The extracted measures are used in some classification methods in order to demonstrate the accuracy of prediction with different DNA regions.
Among the adopted sequences in this work, should be highlighted the TSS dataset, due to the difficulty to recognize them in the genomes of different species,
which many other researches demonstrate techniques for predicting these regions, such as \cite{ohler2000promoter,down2002computational,sonnenburg2006arts}.

The next sections of this paper present in more detail the proposed approach for feature extraction from genomic sequences, 
the adopted genomic sequences datasets, the results and the conclusions.

%
%
%
\vspace{-0.2cm}
\section{Background}\label{sec:background}
\vspace{-0.1cm}
\subsection{Complex Networks}
\vspace{-0.1cm}
\label{ssec:complex networks}
Considered an extension of graph theory, the complex networks is defined as a graph 
which shows an irregular structure of nodes connected by edges \cite{costa2007characterization}.
The complex networks theory extends the formalism of graph theory by adding measures and methods based on real properties of a system \cite{costa2007characterization}.
This theory presents multidisciplinary applications, covering various sciences, such as biology, computer science, physics, 
mathematics and sociology, to cite but a few.
Thus, many real-world systems can be represented by means of complex networks, such as the connection between airports \cite{guimera2004modeling},
the Internet \cite{barabasi2000scale}, Social groups \cite{wasserman1994social}, Neural Networks \cite{costa2005hierarchical,rubinov2010complex} and
biological systems \cite{kauffman1971,jeong2000,albert2005,costa2008,lopes2011gene,lopes2014feature}.

\vspace{-0.2cm}
\subsection{Complex Networks measurements}
The complex networks models have different topologies and well-defined properties which can be used to represent 
biological systems and characterize them in terms of complex networks measures \cite{costa2007characterization}.
Among the possible measures to extract from a complex network, it were adopted the measures listed below.
\begin{enumerate}
\item {{\textbf{Average path length.}}}
The length of the shortest path between two nodes $i$ and $j$, $d_{ij}$, is given by the the length of all 
paths connecting these nodes whose lengths are minimal \cite{watts1998collective}. 
Its determination is important to characterize the internal structure of the networks \cite{boccaletti2006complex}. 
Given a distance matrix D, whose elements $d_{ij}$ represent the value of the shortest path between nodes $i$ and $j$.
The average between the values expressed in matrix D is the lowest average path and it is defined as:
\begin{equation}
 \ell = \frac{1}{N(N - 1)} \sum_{ i\neq j} d_{ij}
\end{equation}

\item {{\textbf{Clustering coefficient.}}}
The clustering coefficient also known as transitivity is a agglomerative measure which represents the probability 
of adjacent nodes for a given node being connected, such as in a social network can be represented as the 
probability between two friends (A and B) have a friend (C) in common. 
Depending on the network topology, the value of transitivity can be different.
Transitivity is defined by the following equation:
\begin{equation}
 C^\omega (i) = \frac{1}{S_i(k_i - 1)} \sum_{j,h} \frac{e_{i,j} + e_{i,j}}{2} a_{i,j}a_{i,h}a_{j,h}
\end{equation}

\item {{\textbf{Centrality.}}}
In the context of complex networks theory, there are different kinds of centrality, such as:
\begin{itemize}
\item Degree Centrality: is defined as the number of links incident on a given node. 
In the case of a directed networks is usual to define two separate measures of degree centrality: \textsl{indegree} e \textsl{outdegree}.

\item Proximity: is the natural distance between a node to all others, i.e., the more central the node is, 
the shorter the distance for all others.

\item Centrality of intermediation: is a measure that quantifies the number of times a node acts as an intermediary on a path between two nodes \cite{freeman1977set}. 

\item Efficiency Centrality: indicates the eccentricity of a node over another, i.e., indicates the fastest way to connect with another node, 
where the smaller its most efficient eccentricity is the node.
\end{itemize}

\item {{\textbf{Average Degree.}}}
The average grade is the arithmetic average of the network degrees, which can be obtained by dividing the number of edges by the number of nodes.

\item {{\textbf{\textsl{Motifs}.}}}
The \textsl{motifs} are subgraphs identified with great frequency within a complex network, which are directly related to the structure and evolution of a 
complex network \cite{milo2002network}.
\item {{\textbf{Number of communities}.}}
Most networks are often modular, i.e., the connections are more frequent between nodes belonging to the same group and less 
frequent between nodes of different groups \cite{danon2005comparing}.
\end{enumerate}

\vspace{-0.1cm}
\subsection{Entropy and Mutual Information}
\label{subsection:entropia}
\vspace{-0.1cm}
The entropy has been defined in the context of information theory by Claude Shannon\cite{shannon2001mathematical}, 
where it indicate the amount of information contained in a particular source, but may also scale the disorder of 
a source of data.

The maximum entropy \cite{jaynes1957} is estimated from the probability distribution by considering its maximum point of uncertainty,
in which the greater the amount of unreliable symbols and patterns in a given source, the greater the entropy of the whole, being defined as:
\begin{equation}
H(X) = - \sum_{x \in X} P(x)logP(x)
\vspace{-0.3cm}
\end{equation}

\subsubsection{Sum and Maximization of Entropy.}
Once the entropy measures the amount of information of a given source, the sum represents the steps for finding the maximum point of the information, 
given by the sum\cite{jaynes1957}:
\begin{equation}
 f(X) = \sum_{i = 1}^{n}P_{i}f(x_{i})
\end{equation}
and the maximization of entropy is defined by the given equation: 
\begin{equation}
 S_{max} = \lambda_{0} + \lambda_{1}(f_{1}(x)) + ... + \lambda_{m}(f_{m}(x))
 \label{eq:maxentropy}
 \vspace{-0.3cm}
\end{equation}
%
%
%
\section{Materials and Methods}
\label{sec:methods}
\vspace{-0.1cm}
\subsection{Genome Sequence Datasets}
\label{ssec:materials}
\vspace{-0.1cm}
\subsubsection{DBTSS.}
The DataBase of Transcriptional Start Sites is a dataset that contains exact positions of transcriptional start sites (TSS), 
determined with the technique called tss-seq in the genomes of various species \cite{yamashita2012dbtss}.
In this work, was adopted a subset containing 1500 promoter sequences randomly chosen. 
The dataset is available at \url{ftp://ftp.hgc.jp/pub/hgc/db/dbtss/Yamashita_NAR/}.
\vspace{-0.3cm}
\subsubsection{Genome Browser.}
We extracted the protein coding  sequences from a set of 1600 randomly selected RefSeq genes from the human genome (hg18). 
To obtain the protein coding sequences, the intronic region of each gene were removed and the coding segments were merged together.  
The sequence of the genes annotated in reverse strand were reverse complemented. 
Using the same genome we also randomly selected 1520  non-coding regions. 
We obtained the annotation of each gene and the hg18 from the UCSC Genome Browser database \cite{kent2002human}.
%
%
%
\vspace{-0.2cm}
\subsection{Feature Extraction Approach}
\label{ssec:methods}
\vspace{-0.1cm}
For the proposed feature extraction were used two sources: complex networks measurements and information theory.
The entropy extracts global features of each sequence, by contrast the complex networks measurements extracts 
local measures of the relationships between nucleotides of the genomic sequence.
\vspace{-0.1cm}
\subsubsection{Information Theory.}
Given a FASTA file, the same processing is performed for each sequence was extracted the following measures: 
histogram, total entropy of the sequence, sum of entropy and maximum value of entropy (See sec. \ref{subsection:entropia}).
For each sequence is considered the nucleotides, dinucleotides and trinucleotides nucleotides, thus, 
thus for each sequence are generated three feature vectors.
\vspace{-0.3cm}
\subsubsection{Complex Networks.}The number of networks is set in accordance with the parameters considered for the method.
In the proposed approach the genomic sequences are used to build complex networks.
The number of generated networks depends of the adopted parameters for the method execution.
Whereas the nodes of the network are represented by occurrence of nucleotides/dinucleotides/trinucleotides within the sequence
the following parameters are needed:
\begin{enumerate}
\item{Word Size ($WS$): is the number of characters that represent the nodes.}
\item{Step ($P$): is the number of characters that will be considered for walking in the sequence after read a word (node) to obtain a new word and constitution of the link between the nodes.}
\end{enumerate}

\begin{figure}[!ht]
\vspace{-0.8cm}
\centering
\subfigure[\label{fig:rededin2x1}]{\includegraphics[width=0.5\linewidth]{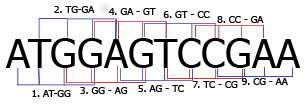}}
\subfigure[\label{fig:rede2x1}]{\includegraphics[width=0.43\linewidth]{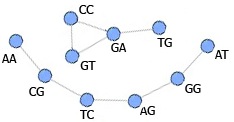}}
\vspace{-0.3cm}
\caption{Application of the proposed methodology for a dinucleotides network by considering the sequence ATGGAGTCCGAA with parameters $P = 1$ and $WS = 2$.
(a) is how the nodes and edges are defined and (b) is the resulting network.}
 \vspace{-0.3cm}
\label{fig:conj}
\end{figure}

In order to better illustrate the proposed approach, an example considering a dinucleotides network, 
with the parameters $WS = 2$ e $P = 1$ is presented in Figure \ref{fig:conj}.
The connection between the nodes is done by considering a $WS = 2$ with the next word in the sequence by considering $P=1$.
Thus, by considering the genomic sequence ATGGAGTCCGAA, the obtained connections with the adopted parameters are presented in Figure \ref{fig:conj}(b).
%
%

By considering the possible of the parameters values WS = \{1,2,3\} and P = \{1,2,3\}, could be generated 6 nondirectional complex networks 
without any nucleotide is disregarded: nucleotides network (WS = 1; P = 1), dinucleotides network (WS = 2; P = 1) and (WS = 2; P = 2), 
trinucleotides (WS = 3; P = 1), (WS = 3; P = 2) and (WS = 3; P = 3).
For each network was extract its complex networks measurements (See sec. \ref{ssec:complex networks}) and some measures of descriptive statistics such as: 
standard deviation, maximum and minimum value of the nodes degree.

The generated feature vector for each genomic sequence is defined by the extracted features of the complex networks and the information theory measures.
%
%
\vspace{-0.4cm}
\section{Results and discussion}\label{sec:results}
\vspace{-0.3cm}
In order to evaluate the proposed feature extraction approach, it was adopted the feature extraction 
and classification from datasets presenting $3$ different DNA regions: 
Coding, Intergenic and hspromoter (See Sec. \ref{ssec:materials}).
It was adopted different classification methods such as \textsl{Naive Bayes}, IBK, 
\textsl{MultiLayer Perceptron}, SVM, J48 and \textsl{Random Forest} 
from WEKA machine learning software \cite{Hall:2009:WDM:1656274.1656278}.

The results were obtained by adopting the Ten-Fold Cross-Validation, by using the default
parameters for the adopted classification methods. 
Only for SVM, the parameter \textsl{Kernel Type} was changed from linear to radial.
All the extracted features were considered in the classification task by all classifiers.
\begin{table}[h]
    \vspace{-0.4cm}
    \centering   
    \caption{Accuracy rating by considering all adopted classifiers methods.}
    \begin{tabular}{c|c}
        \textbf{Classification Method} &  \textbf{Accuracy (\%)} \\
    	\hline
        \textbf{Random Forest} & \textbf{91.2} \\
        SVM   & 84.8 \\
        MultiLayer Perceptron & 64.5\\
        IBK & 79.9\\
        J48 & 89.1\\
        Naive Bayes & 72.6\\
    \end{tabular}
    \label{table:result}
     \vspace{-0.6cm}
\end{table}
%

Table \ref{table:result} shows the rate of correct classification for each adopted classifier.
It is possible to notice that the better results were achieved by the methods \textsl{Random Forest} and \textsl{J48} 
with respectively 91.2\% and 89.1\%, indicating the potential value of this work by using only the genomic sequence itself.
\begin{figure}[!ht]
\vspace{-0.7cm}
\centering
\subfigure[\label{fig:cdsroc}]{\includegraphics[width=0.32\linewidth]{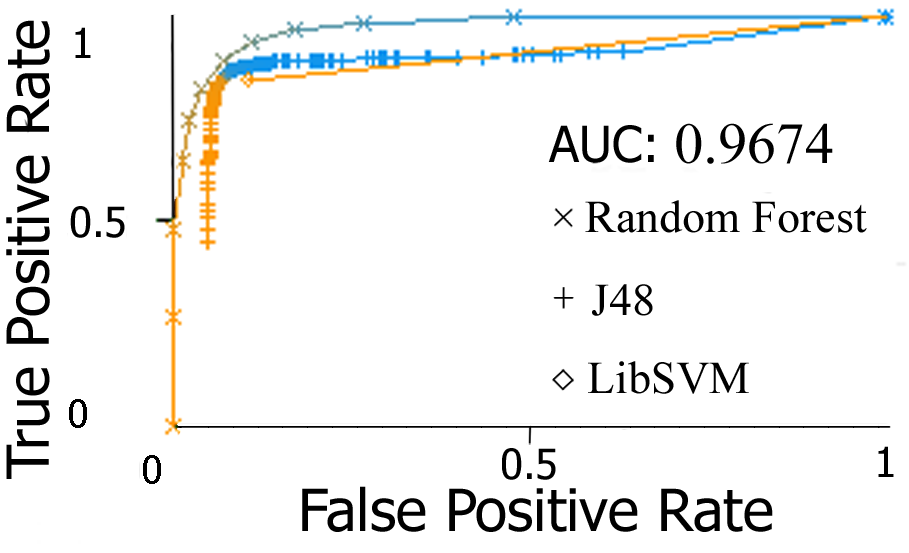}}
\subfigure[\label{fig:interroc}]{\includegraphics[width=0.32\linewidth]{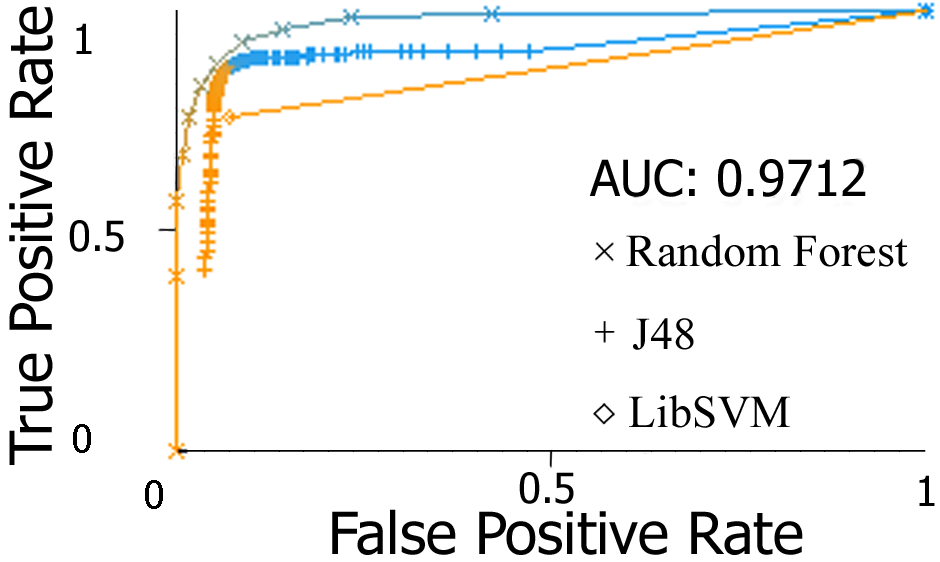}}
\subfigure[\label{fig:promoterroc}]{\includegraphics[width=0.32\linewidth]{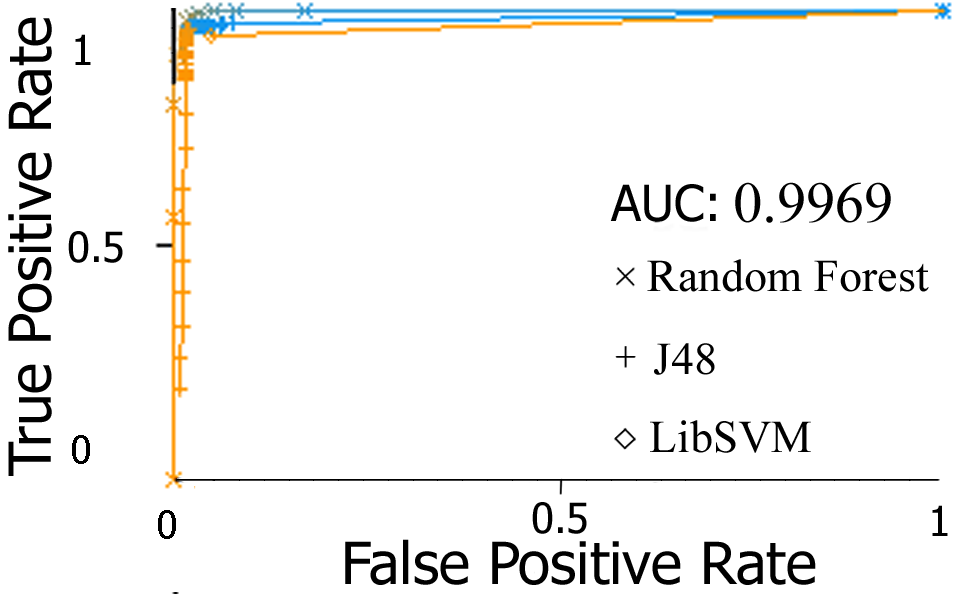}}
\vspace{-0.4cm}
\caption{ROC curves for the Random Forest, J48 e SVM classifiers by considering (a) cds, (b) intergenic and (c) hspromoter genomic sequences.}
 \vspace{-0.5cm}
\label{fig:roc}
\end{figure}
%
%
%
%

Figures \ref{fig:roc} (a), (b) and (c), shows ROC curves for the adopted
classifiers with better performance by considering the classes: Coding, Intergenic and hspromoter.
It is possible to notice the effectiveness of the extracted features by the classification results of 
these three classifiers, indicating the suitability of the proposed approach. 
The characterization of the genomic sequences in both strategies, complex networks and information theory, 
is directly related to the frequency of nucleotides occurrence within them.
Thus, intergenic regions which have a higher repetition of identical nucleotides is detected by the proposed approach,
making possible the identification of theses sequences more correctly and also contributing by the correct inference 
from the other classes.
\vspace{-0.2cm}
\section{Conclusions}\label{sec:conclusion}
\vspace{-0.1cm}
This paper presents a new approach for feature extraction and classification of 
genomic sequences by identifying patterns in the arrangement of the nucleotides 
within them. 
The proposed approach combines complex networks and information theory measures
for the composition of a feature vector representing each genomic sequence.

The experimental results show the suitability and the potential of the proposed 
methodology to classify three different DNA regions: Coding, Intergenic and hspromoter
achieving 91.2\% (\textsl{Random Forest}), 89.1\% (\textsl{J48}) and 84.8\% (\textsl{SVM})
of classification accuracy.

In addition to the obtained results, the complex networks also provides the possibility of 
new experiments because the flexibility of the algorithm to higher parameters (WS and P) 
combination values,  as a result allowing a greater number of networks and extracted features, 
which may improve the classification results. 
Besides, in a further work the proposed methodology can be applied 
for the classification of other classes of genomic sequences such as: miRNA, transposon,
protein-coding genes, RNA genes, regulatory sequences and other sequences.
\vspace{-0.3cm}
\section*{Acknowledgments}
\vspace{-0.3cm}
This work was supported by CNPq and Funda\c{c}\~{a}o Arauc\'{a}ria.
\vspace{-0.3cm}
\bibliographystyle{splncs}
\vspace{-0.1cm}
\bibliography{paper}

\begin{thebibliography}{10}

\bibitem{shen2002network}
Shen-Orr, S.S., Milo, R., Mangan, S., Alon, U.:
\newblock Network motifs in the transcriptional regulation network of
  escherichia coli.
\newblock Nature genetics \textbf{31}(1) (2002)  64--68

\bibitem{diambra2005complex}
Diambra, L., Costa, L.d.F.:
\newblock Complex networks approach to gene expression driven phenotype
  imaging.
\newblock Bioinformatics \textbf{21}(20) (2005)  3846--3851

\bibitem{lopes2014feature}
Lopes, F.M., Martins~Jr, D.C., Barrera, J., Cesar~Jr, R.M.:
\newblock A feature selection technique for inference of graphs from their
  known topological properties: Revealing scale-free gene regulatory networks.
\newblock Information Sciences \textbf{272} (2014)  1--15

\bibitem{lopes2011gene}
Lopes, F.M., Cesar~Jr, R.M., Costa, L.D.F.:
\newblock Gene expression complex networks: synthesis, identification, and
  analysis.
\newblock J. of Comp. Biol. \textbf{18}(10) (2011)  1353--1367

\bibitem{jeong2001lethality}
Jeong, H., Mason, S.P., Barab{\'a}si, A.L., Oltvai, Z.N.:
\newblock Lethality and centrality in protein networks.
\newblock Nature \textbf{411}(6833) (2001)  41--42

\bibitem{costa2006protein}
Costa, L.d.F., Rodrigues, F.A., Travieso, G.:
\newblock Protein domain connectivity and essentiality.
\newblock Applied physics letters \textbf{89}(17) (2006)  174101

\bibitem{jeong2000large}
Jeong, H., Tombor, B., Albert, R., Oltvai, Z.N., Barab{\'a}si, A.L.:
\newblock The large-scale organization of metabolic networks.
\newblock Nature \textbf{407}(6804) (2000)  651--654

\bibitem{vogelstein2000surfing}
Vogelstein, B., Lane, D., Levine, A.J.:
\newblock Surfing the p53 network.
\newblock Nature \textbf{408}(6810) (2000)  307--310

\bibitem{costa2007characterization}
Costa, L.d.F., Rodrigues, F.A., Travieso, G., Villas-Boas, P.R.:
\newblock Characterization of complex networks: a survey of measurements.
\newblock Adv. in Phys. \textbf{56}(1) (2007)  167--242

\bibitem{butte2000}
Butte, A., Kohane, I.:
\newblock Mutual information relevance networks: functional genomic clustering
  using pairwise entropy measurements.
\newblock In: PSB. (2000)  418--429

\bibitem{lopes2008b}
Lopes, F.M., Martins-Jr, D.C., Cesar-Jr, R.M.:
\newblock Feature selection environment for genomic applications.
\newblock BMC Bioinformatics \textbf{9}(1) (October 2008)  451

\bibitem{lopes2011tsallis}
Lopes, F.M., de~Oliveira, E.A., Cesar-Jr, R.M.:
\newblock Inference of gene regulatory networks from time series by {Tsallis}
  entropies.
\newblock BMC Syst. Biology \textbf{5}(1) (2011) ~61

\bibitem{lopes2014b}
Lopes, F.M., Ray, S.S., Hashimoto, R.F., Jr., R.M.C.:
\newblock Entropic biological score: a cell cycle investigation for {GRNs}
  inference.
\newblock Gene \textbf{541}(2) (2014)  129--137

\bibitem{theodoridis2008}
Theodoridis, S., Koutroumbas, K.:
\newblock Pattern Recognition. 4th edn.
\newblock Academic Press, USA (2008)

\bibitem{ohler2000promoter}
Ohler, U.:
\newblock Promoter prediction on a genomic scale - the adh experience.
\newblock Genome research \textbf{10}(4) (2000)  539--542

\bibitem{down2002computational}
Down, T.A., Hubbard, T.J.:
\newblock Computational detection and location of transcription start sites in
  mammalian genomic dna.
\newblock Genome research \textbf{12}(3) (2002)  458--461

\bibitem{sonnenburg2006arts}
Sonnenburg, S., Zien, A., R{\"a}tsch, G.:
\newblock Arts: accurate recognition of transcription starts in human.
\newblock Bioinformatics \textbf{22}(14) (2006)  e472--e480

\bibitem{guimera2004modeling}
Guimera, R., Amaral, L.A.N.:
\newblock Modeling the world-wide airport network.
\newblock The European Physical J. B-Cond. Mat. and Complex Syst.
  \textbf{38}(2) (2004)  381--385

\bibitem{barabasi2000scale}
Barab{\'a}si, A.L., Albert, R., Jeong, H.:
\newblock Scale-free characteristics of random networks: the topology of the
  world-wide web.
\newblock Physica A: Statistical Mechanics and its Applications \textbf{281}(1)
  (2000)  69--77

\bibitem{wasserman1994social}
Wasserman, S.:
\newblock Social network analysis: Methods and applications. Volume~8.
\newblock Cambridge university press (1994)

\bibitem{costa2005hierarchical}
Costa, L.d.F., Sporns, O.:
\newblock Hierarchical features of large-scale cortical connectivity.
\newblock The European Physical J. B-Cond. Mat. and Complex Syst.
  \textbf{48}(4) (2005)  567--573

\bibitem{rubinov2010complex}
Rubinov, M., Sporns, O.:
\newblock Complex network measures of brain connectivity: uses and
  interpretations.
\newblock Neuroimage \textbf{52}(3) (2010)  1059--1069

\bibitem{kauffman1971}
Kauffman, S.A.:
\newblock Chapter 5 gene regulation networks: A theory for their global
  structure and behaviors.
\newblock Volume~6 of Current Topics in Developmental Biology.
\newblock Academic Press (1971)  145--182

\bibitem{jeong2000}
Jeong, H., Tombor, B., Albert, R., Oltvai, Z.N., Barab\'{a}si, A.L.:
\newblock The large-scale organization of metabolic networks.
\newblock Nature \textbf{407} (2000)  651--654

\bibitem{albert2005}
Albert, R.:
\newblock {Scale-free networks in cell biology}.
\newblock J Cell Sci \textbf{118}(21) (2005)  4947--4957

\bibitem{costa2008}
Costa, L.d.F., Rodrigues, F.A., Cristino, A.S.:
\newblock Complex networks: the key to systems biology.
\newblock Genetics and Molecular Biology \textbf{31}(3) (2008)  591--601

\bibitem{watts1998collective}
Watts, D.J., Strogatz, S.H.:
\newblock Collective dynamics of 'small-world' networks.
\newblock nature \textbf{393}(6684) (1998)  440--442

\bibitem{boccaletti2006complex}
Boccaletti, S., Latora, V., Moreno, Y., Chavez, M., Hwang, D.U.:
\newblock Complex networks: Structure and dynamics.
\newblock Physics reports \textbf{424}(4) (2006)  175--308

\bibitem{freeman1977set}
Freeman, L.C.:
\newblock A set of measures of centrality based on betweenness.
\newblock Sociometry (1977)  35--41

\bibitem{milo2002network}
Milo, R., Shen-Orr, S., Itzkovitz, S., Kashtan, N., Chklovskii, D., Alon, U.:
\newblock Network motifs: simple building blocks of complex networks.
\newblock Science \textbf{298} (2002)  824--827

\bibitem{danon2005comparing}
Danon, L., Diaz-Guilera, A., Duch, J., Arenas, A.:
\newblock Comparing community structure identification.
\newblock J. of Statistical Mech.: Theory and Exp. \textbf{2005}(09) (2005)
  P09008

\bibitem{shannon2001mathematical}
Shannon, C.E.:
\newblock A mathematical theory of communication.
\newblock ACM SIGMOBILE Mobile Computing and Communications Review
  \textbf{5}(1) (2001)  3--55

\bibitem{jaynes1957}
Jaynes, E.:
\newblock Information theory and statistical mechanics.
\newblock Physical Review \textbf{106} (1957)  620--630

\bibitem{yamashita2012dbtss}
Yamashita, R., Sugano, S., Suzuki, Y., Nakai, K.:
\newblock Dbtss: Database of transcriptional start sites progress report in
  2012.
\newblock NAR \textbf{40}(D1) (2012)  D150--D154

\bibitem{kent2002human}
Kent, W.J., Sugnet, C.W., Furey, T.S., Roskin, K.M., Pringle, T.H., Zahler,
  A.M., Haussler, D.:
\newblock The human genome browser at {UCSC}.
\newblock Genome Research \textbf{12}(6) (2002)  996--1006

\bibitem{Hall:2009:WDM:1656274.1656278}
Hall, M., Frank, E., Holmes, G., Pfahringer, B., Reutemann, P., Witten, I.H.:
\newblock The weka data mining software: An update.
\newblock SIGKDD Exp. Newsl. \textbf{11}(1) (2009)  10--18

\end{thebibliography}
\vspace{-0.5cm}
\end{document}